# A Completely Blind Channel Estimation Technique for OFDM Using Constellation Splitting

Sameera Bharadwaja H.

*ASE in C1 grade (Researcher), CTO org.
TCS Innovation labs, Delhi (Gurgaon branch)
E-mail id: sameerabharadwaja@gmail.com*

D. K. Mehra

*Professor, Dept. of E&C Engg.,
Indian Institute of Technology Roorkee, Roorkee
E-mail id: dkmecfec@iitr.ernet.in*

**Abstract**

*The problem of second-order statistics (SOS)-based blind channel estimation in OFDM systems is addressed in this paper. Almost all SOS-based methods proposed so far suffer from a complex-scalar estimation ambiguity, which is resolved by using pilots or reference symbols. We propose an algorithm to resolve this ambiguity in blind manner using frequency-domain linear non-redundant precoding and constellation-splitting among the alternate subcarriers. The performance of the proposed scheme is evaluated via numerical simulations in MATLAB environment. Simulation results show that the proposed approach performs as good as its semi-blind counterpart for M-ary PAM systems.*

## I. INTRODUCTION

In a digital communication link, prior to the coherent detector block, the incoming information symbol needs to be equalized of imperfect channel variations to prevent erroneous decisions [1]. The channel impulse response (CIR) knowledge is crucial to ensure acceptable performance of equalizer. In case of non-coherent detection (for differentially coded data) equalization is not compulsory, but use of non-coherent detection causes SNR loss of around 3-dB. In practice, the CIR is estimated by using training. But, for time-varying channels, training based methods are unsuitable because of high bandwidth overhead introduced [2]. The blind techniques (no training) have been researched during past three decades [3, 4, 5 and references therein summarises almost all blind approaches proposed up to last decade]. Since training overhead is eliminated; higher spectral efficiency and higher information rates can be achieved.

The early blind approaches are based on utilization of higher-order statistics (HOS). These methods are useful in situations where-in the non-Gaussianity, non-minimum phase, coloured noise or non-linearity have to be accounted for. Extracting information from HOS involves non-linear processing (computationally complex) and has low convergence rate. This compromise on convergence rate is unacceptable in applications where-in the channel is time-varying (mobile channels), data rates are high or the data is sent in short bursts (TDMA).

In mid 1990's and onwards, the SOS-based approaches were proposed for single-carrier systems. These methods have relatively low processing demands and good convergence rates. SOS contains enough information to estimate amplitude response of the channel only, which restricts it to minimum phase systems. Further most of SOS-based approaches fail in estimation of arbitrary (critical and singular) channels.

In the early part of past decade, multi-carrier transmission, especially OFDM have gained much importance. OFDM converts frequency-selective channel to flat-fading channel and with cyclic-prefix, channel estimation and equalization can be achieved in frequency-domain (one-tap complex equalizer per subcarrier) [6]. Further, OFDM has been adopted as a standard in DAB, DVB (HDTV), Wireless LAN (IEEE 802.11), WiMAX (IEEE 802.16), 3 GPP Long-Term Evolution (LTE) etc. Over time, blind SOS-based OFDM channel estimation methods have been proposed [7, 8, 9]. Recently, precoding technique at the transmitter has been adopted to introduce certain correlation among the subcarriers in OFDM frame [10, 11]. This correlation information is used to estimate channel up to a constant complex-scalar ambiguity. Further, by precoding, estimation of arbitrary channels using SOS is possible, thereby making SOS methods attractive over HOS-based methods.

The phase ambiguity fundamental to most of the SOS-based methods are resolved in [7-11] via pilot/ reference symbols. By using pilots, the charm of blind methods is lost. According to the author's knowledge, very little work on resolving this ambiguity blindly using SOS-based methods has been done [12, 13 and 14]. The authors of [12] have restricted their work to M-ary PSK which has constant modulus property and use variable-rate source coding (not practically feasible) and adopts ML estimator. ML estimator demands high implementation complexity. Further, QAM has been used extensively as a modulation scheme in most of wireless telecommunication standards (IEEE 802.11 series). The algorithm proposed in [14] use asymmetric 4-QAM constellation with non-zero average energy (dc-level) which suffers from constant wastage of power.

This paper proposes a blind algorithm to resolve the ambiguity by using received signal SOS in blind manner (no pilots/ reference symbols/ training) via frequency-domain linear non-redundant precoding [11] and constellation-splitting among the alternate subcarriers for M-ary PAM systems. The rest of the paper is organized as follows: Section II presents the system model. Section III presents the proposed algorithm. In Section IV, the simulations results are provided to show that the proposed method performs as good as its semi-blind counterpart. Conclusions and scope for future work in this area are discussed in Section V. Standard mathematical notations and MATLAB notations are used.

## II. SYSTEM MODEL

Figure 1 shows a model for the baseband SISO-OFDM system. The OFDM transceiver block is similar to one described in [6]. Each OFDM frame consists of M subcarriers. The channel is assumed to be frequency-selective represented by vector $\mathbf{h} = [h_0, \ldots, h_L]^T$.

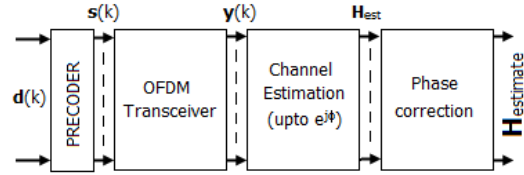

*Figure 1*: Baseband precoded SISO-OFDM system model with channel estimation

Let $\mathbf{H} = [H_0, \ldots, H_{M-1}]^T = DFT(\mathbf{h})$ be the channel frequency response. Assuming that sufficient CP is used, the frequency-domain received signal vector for the $k$th block is given by,

$$y(k) = \tilde{\mathbf{H}}s(k) + n(k) \qquad (1)$$

where, $\tilde{\mathbf{H}} = diag\{\mathbf{H}\}$ is the diagonal channel matrix; $n(k)$ is $M \times 1$ additive white Gaussian noise vector; $s(k)$ is $M \times 1$ precoded frequency-domain transmitted vector and $y(k)$ is $M \times 1$ received vector in frequency-domain. The symbol vector $d(k)$ is precoded by a matrix $\mathbf{W}$ to obtain $s(k)$. Thus,

$$y(k) = \tilde{\mathbf{H}}\mathbf{W}d(k) + n(k) \qquad (2)$$

Hence, the signal covariance matrix is,

$$R = \tilde{H}WR_dW^H\tilde{H}^H + \sigma_n^2 I \quad (3)$$

Assuming that the source covariance matrix, $R_d = E\{d(k)d(k)^H\} = \sigma_d^2 I$ and $P = WW^H$. After some mathematical modifications (3) can be rewritten as [11],

$$R = \sigma_d^2(HH^H) \odot P + \sigma_n^2 I \quad (4)$$

where, $\odot$ represents Hadamard product. It is required to estimate the vector *H* (or *h*) using the knowledge of correlation information in *R* (introduced due to precoding at the transmitter [11]) as shown in (4). The statistical knowledge of *d(k)* and the phase relationship information among alternate subcarriers in transmitted frame is assumed to be known at the receiver.

## III. PROPOSED ALGORITHM

The channel response is estimated up to a constant phase ambiguity using joint estimation algorithm of [11]. The signal covariance matrix is calculated by time-averaging. The correlation information in all the columns is used via joint algorithm of [11] to estimate the channel vector in frequency-domain. This estimate is of the form $H_{est} = H\ e^{j\varphi}$ where, **H** is the true channel response vector in frequency-domain; $H_{est} = [\hat{H}_0 \cdots \hat{H}_{M-1}]^T$ is the first approximation of **H** and $\varphi$ is the phase ambiguity. The phase correction algorithm for SISO-OFDM with PAM mapping is discussed. Consider an OFDM frame, with each of the alternate subcarrier symbol chosen from an (M/2)-ary PAM constellation instead of M-ary constellation. For even numbered subcarriers, the constellation points are shifted to right by (M/2) such that the symbols have zero phases and for odd numbered subcarriers, they are shifted to left by (M/2) such that they have a phase of $180^0$ each. This is analogous to splitting the M-ary constellation along the axis of symmetry into two subsets; assigning the right subset to even numbered subcarriers and the left subset to odd numbered subcarriers. This is known as 'Constellation-splitting'. The concept is illustrated in Fig. 2 for PAM system with M = 8. Single OFDM frame thus consist of symbols chosen from M-ary constellation and have zero dc in statistical sense. Also, a unique phase relationship is introduced among the alternate subcarriers. This information can be used at the receiver for phase-ambiguity estimation and correction. Consider,

$$d = [|d_0|e^{j0} \quad |d_1|e^{j\pi} \quad \cdots \quad |d_{M-1}|e^{j\pi}]^T$$
... (5)

where,

$$d_i \in \begin{cases} Right\ signal\ subset, & i\ even \\ Left\ signal\ subset, & i\ odd \end{cases}$$

The precoder matrix **W** is chosen such that the phase relationship among the alternate subcarriers is retained. Thus,

$$s = Wd = [|s_0|e^{j\theta} \quad \cdots \quad |s_{M-1}|e^{j(\theta+\pi)}]^T$$
... (6)

Assuming sufficient CP, the received signal vector is given by,

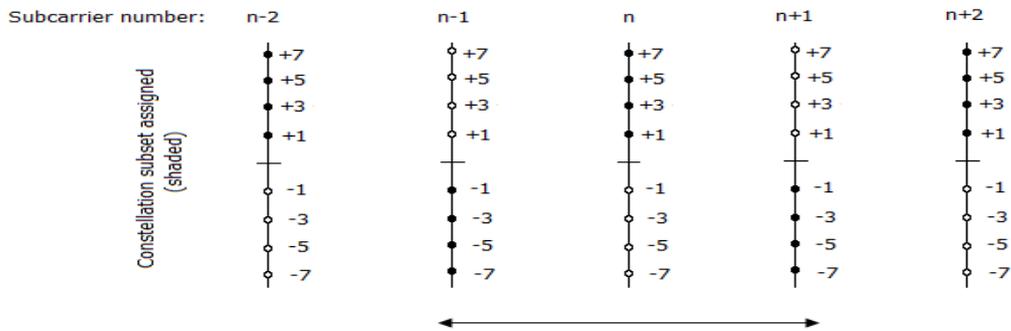

*Figure 2*: *Constellation subset assignment scheme among alternate subcarriers in an OFDM frame for 8-PAM mapping*

$$\mathbf{y} = \begin{bmatrix} |s_0||H_0|e^{j(\theta+\theta_1)} \\ \vdots \\ |s_{M-1}||H_{M-1}|e^{j(\theta+\pi+\theta_{M-1})} \end{bmatrix} \quad (7)$$

where, **H** is the frequency-domain channel vector given by,

$$\mathbf{H} = \begin{bmatrix} |H_0|e^{j\theta_1} & \cdots & |H_{M-1}|e^{j\theta_{M-1}} \end{bmatrix}^T$$

The first-order estimate of **H** up to a constant phase ambiguity as obtained from joint estimation algorithm of [11] can be expanded as,

$$\mathbf{H}_{est}^T = \begin{bmatrix} |H_{0e}|e^{j(\theta_{1e}+\varphi)} & \cdots & |H_{(M-1)e}|e^{j(\theta_{(M-1)e}+\varphi)} \end{bmatrix}$$

... (8)

where, $|H_{ie}|$ is the estimate of $|H_i|$; $\theta_{ie}$ is the estimate of $\theta_i$ and $\varphi$ is the phase ambiguity. From (7) and (8), it is clear that the phase ambiguity per subcarrier can be obtained as (with $\theta = 0$),

$$\varphi_{i(est)} = Ang\{\mathbf{H}_{est}\} - Ang\{y(i)\} + \mathbf{B}$$

for $i = 0$ to $M-1$

... (9)

where, $\mathbf{B} = [0, 180^0, 0, \ldots, 180^0]^T$. The value of $\varphi_{i(est)}$ is averaged over sufficient number of OFDM frames to combat the ill-effects of AWGN. The final estimate of ambiguity term: $\varphi_{est}$ is given by,

$$\varphi_{est} = Mean\{\varphi_{i(est)}\} \quad (10)$$

Using the result of (10) in (8), the phase error in $\mathbf{H}_{est}$ can be corrected to obtain $\mathbf{H}_{estimate}$

$$Ang\{\mathbf{H}_{estimate}(i)\} = Ang\{\mathbf{H}_{est}(i)\} - \varphi_{est}$$

for $i = 0$ to $M-1$ ... (11)

Since no pilots/ reference symbols are used, our method is completely blind.

## IV. SIMULATION RESULTS

The performance of the proposed algorithm is evaluated for a three-tap ($L = 2$) channel with [15]:

a. Exponential delay profile-
$$E\{|h_l|^2\} = e^{(-\frac{l}{10})}, \quad l = 0, \ldots, L$$

b. Uniform delay profile-
$$E\{|h_l|^2\} = 1, \quad l = 0, \ldots, L$$

The channel phase is assumed to be uniformly distributed over [0, 2π). Number of subcarriers per OFDM frame is M = 64. The simulation results are compared, with 4-, 8- and 16-PAM systems. The results are averaged over 100 Monte-Carlo runs. The number of OFDM blocks used is 500 and SNR level is 30 dB unless otherwise mentioned. The precoder matrix **W** used is given by:

$$\mathbf{W} = \begin{bmatrix} 1 & -p & p & \cdots & -p \\ -p & 1 & -p & \cdots & p \\ \vdots & \vdots & \vdots & \ddots & \vdots \\ -p & p & -p & \cdots & 1 \end{bmatrix}_{M \times M}$$

where, $0 < p < 1$. The results are shown for $p = 0.5$.

The simulation results of Fig. 3 show that the proposed estimator performs as good as its semi-blind counterpart [11] with M-ary PAM mapping for a three-tap channel with above mentioned power delay profiles in statistical sense and is independent of the value of M. The semi-blind joint-estimator of [11] has comparatively better performance at lower SNR (< 15 dB) but at high SNR values, the performances of the two are identical. Further, the semi-blind estimator converges faster than the proposed blind estimation algorithm as the number of OFDM blocks used is varied but meets asymptotically.

The proposed phase-correction scheme suffers from a rate loss of 1-bit per OFDM symbol irrespective of the value of M used. This loss can be made insignificant for all practical purposes if the constellation size is kept sufficiently large by choosing higher value of M. The advantage of the proposed method is that it is completely blind.

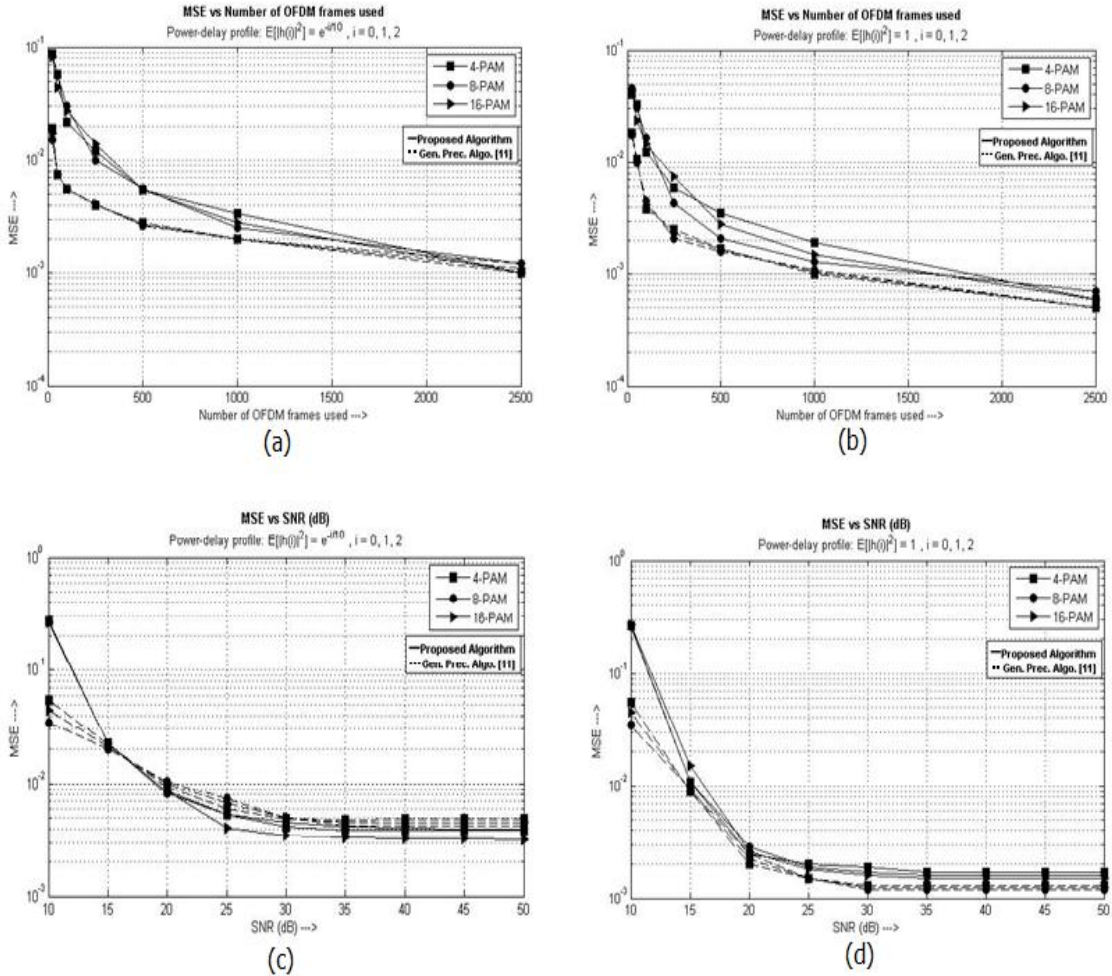

*Figure 3*: Simulation results depicting the performance of proposed blind channel estimator

*(a) – A Plot of MSE vs. Number of OFDM blocks used for a channel with exp. PDP; (b) – A Plot of MSE vs. Number of OFDM blocks used for a channel with uniform PDP; (c) – A Plot of MSE vs. SNR (dB) value for a channel with exp. PDP; (d) – A Plot of MSE vs. SNR (dB) for a channel with uniform PDP*

## V. CONCLUSION

A completely blind channel estimation algorithm based on SOS was proposed in this paper. The simulation results show that, the proposed method performs as good as its semi-blind counterpart [11]. Future work aims at possible extension of proposed method to more general and widely used modulation schemes such as QAM which is nothing but two PAM systems in quadrature and PSK (a degenerative scenario of QAM) and in MIMO-OFDM systems.

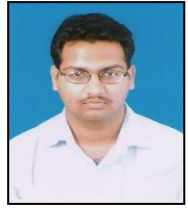
**Mr. Sameera Bharadwaja H** received the M. Tech. degree in Communication Systems from Indian Institute of Technology Roorkee, Uttarakhand, India in 2011; the B.E. degree in Telecommunication Engineering from P.E.S. Institute of Technology, Bangalore, India (affiliated to Visveswaraiah Technological University – Belgaum, Karnataka, India) in 2009 and is currently working as a researcher in Tata Innovation Labs (iLabs), Delhi (Gurgaon branch). His research interests are in wireless communications, signal processing for communication, information/ communication theory and image processing.

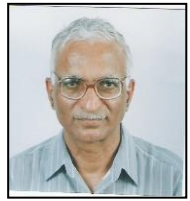
**Dr. D. K. Mehra** received the B.E. and M.E. degree in electrical communication engineering from the Indian Institute of Science, Bangalore, India in 1968 and 1970, respectively, and the Ph.D. degree from the Indian Institute of Technology, Kanpur, India in 1978. In 1975 he joined the Electronics and Computer Engineering Department of Indian Institute of Technology Roorkee (Previously University of Roorkee), where he became a Professor in 1987. He retired in June 2011 and joined as an Emeritus Fellow in the department from July 2011. His main teaching and research interests are in the area of information and communication theory, coding theory, adaptive signal processing techniques, digital communication over fading dispersive channels, channel estimation and interference suppression in CDMA/OFDM systems. He has supervised a number of students for their B. Tech. final year project, M. Tech. dissertation and Ph. D. dissertation in the above areas.